\documentclass[aps,prl,twocolumn,showpacs,superscriptaddress,groupedaddress,floatfix]{revtex4}  
\usepackage{graphicx}  
\usepackage{dcolumn}   
\usepackage{bm}        
\usepackage{amssymb}   
\usepackage{color}

\def\BR{{\cal B}}
\def\ppbar{{p}\overline{{p}}}

\def\Hpm{ {H}^{\pm\pm}}

\def\HpmL{ {H}^{\pm\pm}_{L}}
\def\HpmR{ {H}^{\pm\pm}_{R}}
\def\HppL{ {H}^{++}_{L}}
\def\HmmL{ {H}^{--}_{L}}
\def\HppR{ {H}^{++}_{R}}
\def\HmmR{ {H}^{--}_{R}}
\def\Hpp{ {H}^{++}}
\def\Hmm{ {H}^{--}}
\def\MET{{p\kern -0.4em/_{T}}}

\begin{document}

\hspace{5.2in} \mbox{Fermilab-Pub-11/286-E}

\title{Search for doubly-charged Higgs boson pair production in 
$\boldsymbol\ppbar$ collisions at $\boldsymbol{\sqrt{s}=1.96}$~TeV}
%
\affiliation{Universidad de Buenos Aires, Buenos Aires, Argentina}
\affiliation{LAFEX, Centro Brasileiro de Pesquisas F{\'\i}sicas, Rio de Janeiro, Brazil}
\affiliation{Universidade do Estado do Rio de Janeiro, Rio de Janeiro, Brazil}
\affiliation{Universidade Federal do ABC, Santo Andr\'e, Brazil}
\affiliation{Instituto de F\'{\i}sica Te\'orica, Universidade Estadual Paulista, S\~ao Paulo, Brazil}
\affiliation{Simon Fraser University, Vancouver, British Columbia, and York University, Toronto, Ontario, Canada}
\affiliation{University of Science and Technology of China, Hefei, People's Republic of China}
\affiliation{Universidad de los Andes, Bogot\'{a}, Colombia}
\affiliation{Charles University, Faculty of Mathematics and Physics, Center for Particle Physics, Prague, Czech Republic}
\affiliation{Czech Technical University in Prague, Prague, Czech Republic}
\affiliation{Center for Particle Physics, Institute of Physics, Academy of Sciences of the Czech Republic, Prague, Czech Republic}
\affiliation{Universidad San Francisco de Quito, Quito, Ecuador}
\affiliation{LPC, Universit\'e Blaise Pascal, CNRS/IN2P3, Clermont, France}
\affiliation{LPSC, Universit\'e Joseph Fourier Grenoble 1, CNRS/IN2P3, Institut National Polytechnique de Grenoble, Grenoble, France}
\affiliation{CPPM, Aix-Marseille Universit\'e, CNRS/IN2P3, Marseille, France}
\affiliation{LAL, Universit\'e Paris-Sud, CNRS/IN2P3, Orsay, France}
\affiliation{LPNHE, Universit\'es Paris VI and VII, CNRS/IN2P3, Paris, France}
\affiliation{CEA, Irfu, SPP, Saclay, France}
\affiliation{IPHC, Universit\'e de Strasbourg, CNRS/IN2P3, Strasbourg, France}
\affiliation{IPNL, Universit\'e Lyon 1, CNRS/IN2P3, Villeurbanne, France and Universit\'e de Lyon, Lyon, France}
\affiliation{III. Physikalisches Institut A, RWTH Aachen University, Aachen, Germany}
\affiliation{Physikalisches Institut, Universit{\"a}t Freiburg, Freiburg, Germany}
\affiliation{II. Physikalisches Institut, Georg-August-Universit{\"a}t G\"ottingen, G\"ottingen, Germany}
\affiliation{Institut f{\"u}r Physik, Universit{\"a}t Mainz, Mainz, Germany}
\affiliation{Ludwig-Maximilians-Universit{\"a}t M{\"u}nchen, M{\"u}nchen, Germany}
\affiliation{Fachbereich Physik, Bergische Universit{\"a}t Wuppertal, Wuppertal, Germany}
\affiliation{Panjab University, Chandigarh, India}
\affiliation{Delhi University, Delhi, India}
\affiliation{Tata Institute of Fundamental Research, Mumbai, India}
\affiliation{University College Dublin, Dublin, Ireland}
\affiliation{Korea Detector Laboratory, Korea University, Seoul, Korea}
\affiliation{CINVESTAV, Mexico City, Mexico}
\affiliation{Nikhef, Science Park, Amsterdam, the Netherlands}
\affiliation{Radboud University Nijmegen, Nijmegen, the Netherlands and Nikhef, Science Park, Amsterdam, the Netherlands}
\affiliation{Joint Institute for Nuclear Research, Dubna, Russia}
\affiliation{Institute for Theoretical and Experimental Physics, Moscow, Russia}
\affiliation{Moscow State University, Moscow, Russia}
\affiliation{Institute for High Energy Physics, Protvino, Russia}
\affiliation{Petersburg Nuclear Physics Institute, St. Petersburg, Russia}
\affiliation{Instituci\'{o} Catalana de Recerca i Estudis Avan\c{c}ats (ICREA) and Institut de F\'{i}sica d'Altes Energies (IFAE), Barcelona, Spain}
\affiliation{Stockholm University, Stockholm and Uppsala University, Uppsala, Sweden}
\affiliation{Lancaster University, Lancaster LA1 4YB, United Kingdom}
\affiliation{Imperial College London, London SW7 2AZ, United Kingdom}
\affiliation{The University of Manchester, Manchester M13 9PL, United Kingdom}
\affiliation{University of Arizona, Tucson, Arizona 85721, USA}
\affiliation{University of California Riverside, Riverside, California 92521, USA}
\affiliation{Florida State University, Tallahassee, Florida 32306, USA}
\affiliation{Fermi National Accelerator Laboratory, Batavia, Illinois 60510, USA}
\affiliation{University of Illinois at Chicago, Chicago, Illinois 60607, USA}
\affiliation{Northern Illinois University, DeKalb, Illinois 60115, USA}
\affiliation{Northwestern University, Evanston, Illinois 60208, USA}
\affiliation{Indiana University, Bloomington, Indiana 47405, USA}
\affiliation{Purdue University Calumet, Hammond, Indiana 46323, USA}
\affiliation{University of Notre Dame, Notre Dame, Indiana 46556, USA}
\affiliation{Iowa State University, Ames, Iowa 50011, USA}
\affiliation{University of Kansas, Lawrence, Kansas 66045, USA}
\affiliation{Kansas State University, Manhattan, Kansas 66506, USA}
\affiliation{Louisiana Tech University, Ruston, Louisiana 71272, USA}
\affiliation{Boston University, Boston, Massachusetts 02215, USA}
\affiliation{Northeastern University, Boston, Massachusetts 02115, USA}
\affiliation{University of Michigan, Ann Arbor, Michigan 48109, USA}
\affiliation{Michigan State University, East Lansing, Michigan 48824, USA}
\affiliation{University of Mississippi, University, Mississippi 38677, USA}
\affiliation{University of Nebraska, Lincoln, Nebraska 68588, USA}
\affiliation{Rutgers University, Piscataway, New Jersey 08855, USA}
\affiliation{Princeton University, Princeton, New Jersey 08544, USA}
\affiliation{State University of New York, Buffalo, New York 14260, USA}
\affiliation{Columbia University, New York, New York 10027, USA}
\affiliation{University of Rochester, Rochester, New York 14627, USA}
\affiliation{State University of New York, Stony Brook, New York 11794, USA}
\affiliation{Brookhaven National Laboratory, Upton, New York 11973, USA}
\affiliation{Langston University, Langston, Oklahoma 73050, USA}
\affiliation{University of Oklahoma, Norman, Oklahoma 73019, USA}
\affiliation{Oklahoma State University, Stillwater, Oklahoma 74078, USA}
\affiliation{Brown University, Providence, Rhode Island 02912, USA}
\affiliation{University of Texas, Arlington, Texas 76019, USA}
\affiliation{Southern Methodist University, Dallas, Texas 75275, USA}
\affiliation{Rice University, Houston, Texas 77005, USA}
\affiliation{University of Virginia, Charlottesville, Virginia 22901, USA}
\affiliation{University of Washington, Seattle, Washington 98195, USA}
\author{V.M.~Abazov} \affiliation{Joint Institute for Nuclear Research, Dubna, Russia}
\author{B.~Abbott} \affiliation{University of Oklahoma, Norman, Oklahoma 73019, USA}
\author{B.S.~Acharya} \affiliation{Tata Institute of Fundamental Research, Mumbai, India}
\author{M.~Adams} \affiliation{University of Illinois at Chicago, Chicago, Illinois 60607, USA}
\author{T.~Adams} \affiliation{Florida State University, Tallahassee, Florida 32306, USA}
\author{G.D.~Alexeev} \affiliation{Joint Institute for Nuclear Research, Dubna, Russia}
\author{G.~Alkhazov} \affiliation{Petersburg Nuclear Physics Institute, St. Petersburg, Russia}
\author{A.~Alton$^{a}$} \affiliation{University of Michigan, Ann Arbor, Michigan 48109, USA}
\author{G.~Alverson} \affiliation{Northeastern University, Boston, Massachusetts 02115, USA}
\author{G.A.~Alves} \affiliation{LAFEX, Centro Brasileiro de Pesquisas F{\'\i}sicas, Rio de Janeiro, Brazil}
% ...
%
%
\collaboration{The D0 Collaboration\footnote{with visitors from
$^{a}$Augustana College, Sioux Falls, SD, USA,
$^{b}$The University of Liverpool, Liverpool, UK,
$^{c}$SLAC, Menlo Park, CA, USA,
$^{d}$University College London, London, UK,
$^{e}$Centro de Investigacion en Computacion - IPN, Mexico City, Mexico,
$^{f}$ECFM, Universidad Autonoma de Sinaloa, Culiac\'an, Mexico,
and 
$^{g}$Universit{\"a}t Bern, Bern, Switzerland.
$^{\ddag}$Deceased.
}} \noaffiliation
\vskip 0.25cm
       
\date{\today}

\begin{abstract}
We present a search for pair production of doubly-charged Higgs bosons in the processes
$q\bar{q} \to \Hpp\Hmm$ decaying through
 $\Hpm \to \tau^{\pm}\tau^{\pm},\mu^{\pm}\tau^{\pm},\mu^{\pm}\mu^{\pm}$.
The search is performed in $\ppbar$ collisions
at a center-of-mass energy of $\sqrt{s}=1.96$~TeV using an 
integrated luminosity of up to  $7.0$~fb$^{-1}$ 
collected by the D0 experiment at the Fermilab Tevatron Collider.   The results are used
to set $95\%$ C.L.~limits on the pair production cross section of  doubly-charged Higgs bosons and on their
mass for different $\Hpm$ branching fractions. 
Models predicting different $\Hpm$ decays are investigated. Assuming $\BR(\Hpm\to\tau^{\pm}\tau^{\pm})=1$ yields 
 an observed (expected) lower limit on the mass of a left-handed $\Hpm_L$ boson 
 of 128 (116)~GeV and assuming $\BR(\Hpm\to\mu^{\pm}\tau^{\pm})=1$ the
corresponding limits are 144 (149)~GeV.
In a model with $\BR(\Hpm \to \tau^{\pm}\tau^{\pm})=\BR(\Hpm\to\mu^{\pm}\tau^{\pm})=\BR(\Hpm\to\mu^{\pm}\mu^{\pm})=1/3$,
we obtain $M(\Hpm_L)>130~(138)$~GeV.
\end{abstract}

\maketitle

Doubly-charged Higgs bosons ($\Hpm$) appear in models with an extended Higgs sector such as the Little Higgs model~\cite{bib-littleH}, 
left-right symmetric models~\cite{bib-LRsym}, and
in models with {\it SU$(3)_c\times$SU$(3)_L\times$U$(1)_Y$} (3-3-1) gauge 
symmetry~\cite{bib-331}. 

The $\Hpm$ bosons could be pair-produced and observed at a hadron collider through the process
$q\bar{q} \rightarrow Z/\gamma^* \rightarrow H^{++} H^{--}\to{\ell}^{+}{\ell'}^{+}{\ell}^{-}{\ell'}^{-}$
($\ell,\ell'=e,\mu,\tau$). 
Single production of $\Hpm$ bosons through $W$ exchange, leading to $\Hpm H^{\mp}$ final states, is not considered in this Letter
to reduce the model dependency of the results~\cite{bib-single}.
Some models favor a mass of the $\Hpm$ boson at the electroweak scale~\cite{bib-mass}.
 The decay into like-charge lepton pairs
violates lepton flavor number conservation. 
The decays $\Hpm\to\tau^{\pm}\tau^{\pm}$ are predicted to dominate
in some scenarios, such as the 3-3-1 model of Ref.~\cite{bib-scalar}.
In a Higgs triplet model that is based
on a seesaw neutrino mass mechanism,
a normal hierarchy of neutrino masses 
leads to approximately equal branching fractions for $\Hpm$ boson decays to 
$\tau\tau$, 
$\mu\tau$, and $\mu\mu$, if the mass of the lightest neutrino is less than 
$10$~meV~\cite{bib-seesaw}.  
In this Letter, we present the first comparison of data with this model and
the first search 
for $\Hpm\to\tau^{\pm}\tau^{\pm}$ decays at a hadron collider. 
 
In left-right symmetric models, right-handed states ($\HpmR$) appear in
addition to left-handed states ($\HpmL$). They are characterized through their
coupling to right-handed and left-handed fermions, respectively.
The cross section for production of right-handed $\HppR\HmmR$ pairs is about a factor of 2 smaller
than for $\HppL\HmmL$ because of the different coupling to the $Z$ boson~\cite{bib-spira}.
The mass limits for $\HpmR$ bosons therefore tend to be weaker than for $\HpmL$
bosons.

Searches for production of $\Hpm$ bosons have been performed previously 
at the CERN $e^+e^-$ Collider (LEP)~\cite{bib-lep} and
at the DESY $ep$ Collider (HERA)~\cite{bib-hera}.  
Limits on the mass of the $\Hpm$ boson were obtained 
in the range of $95-100$~GeV, 
depending on the flavor of the final state leptons. 
The OPAL and H1 Collaborations searched for single $\Hpm$ production
in the processes $e^+e^-\to e^{\mp}e^{\mp}\Hpm$~\cite{bib-opal} and 
$e^{\pm}p\to {\ell}^{\mp}\Hpm p$~\cite{bib-hera}, and through the
study of Bhabha scattering $e^+e^-\to  e^+e^-$~\cite{bib-opal},
constraining the $\Hpm$ boson's Yukawa couplings $h_{ee}$ to electrons.
 Bounds on decays such as $\tau\to 3 \mu$ or $\mu\to e\gamma$ and 
the measured $(g-2)_\mu$ also constrain
different $h_{\ell\ell'}$~\cite{bib-moha1}.
At the Fermilab Tevatron Collider, the D0 and CDF Collaborations
published limits for $\mu\mu$, $ee$, $e\tau$, and $\mu\tau$ final states
in the range $M(\HpmL)>112-150$~GeV, assuming $100\%$ decays into the specified final state
~\cite{bib-d0hmm1, bib-d0hmm2, bib-cdf1, bib-cdf2}.


The results in this Letter are based on data collected 
with the D0 detector at the Fermilab Tevatron Collider
and correspond to an integrated luminosity of up to $7.0$~fb$^{-1}$. 
The D0 detector~\cite{d0det} comprises tracking detectors and calorimeters.
Silicon microstrip detectors and a
scintillating fiber tracker are used to reconstruct charged particle tracks 
within a $2$~T solenoid. The
uranium and liquid-argon calorimeters used to measure particle energies
consist of electromagnetic
(EM) and hadronic sections. 
Muons are identified
by combining tracks in the central tracker with patterns of hits in the muon spectrometer.
Events are required to pass triggers that select at least one
muon candidate. 

\begin{figure*}[ht]
\includegraphics[scale=0.29]{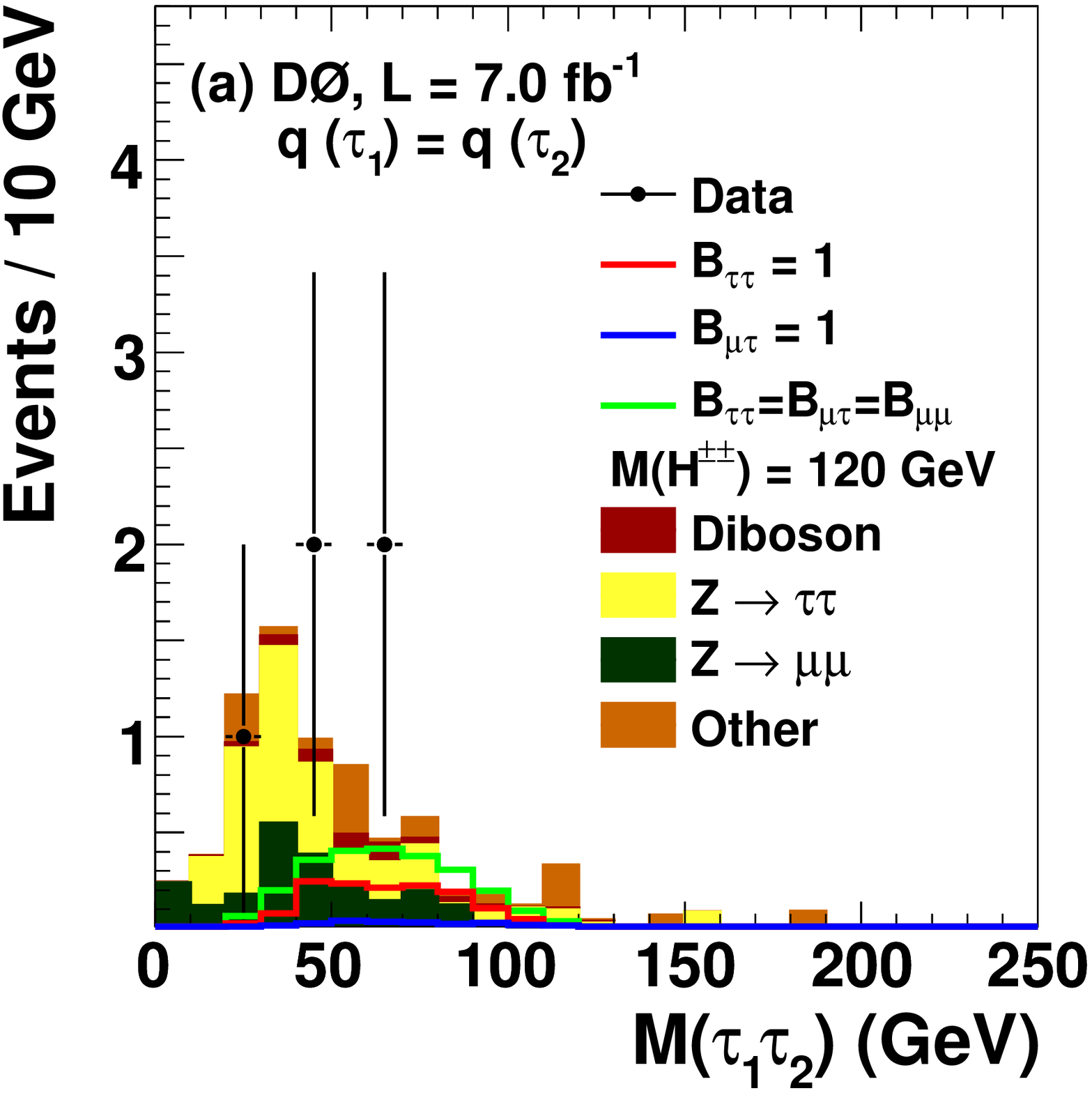}
\includegraphics[scale=0.29]{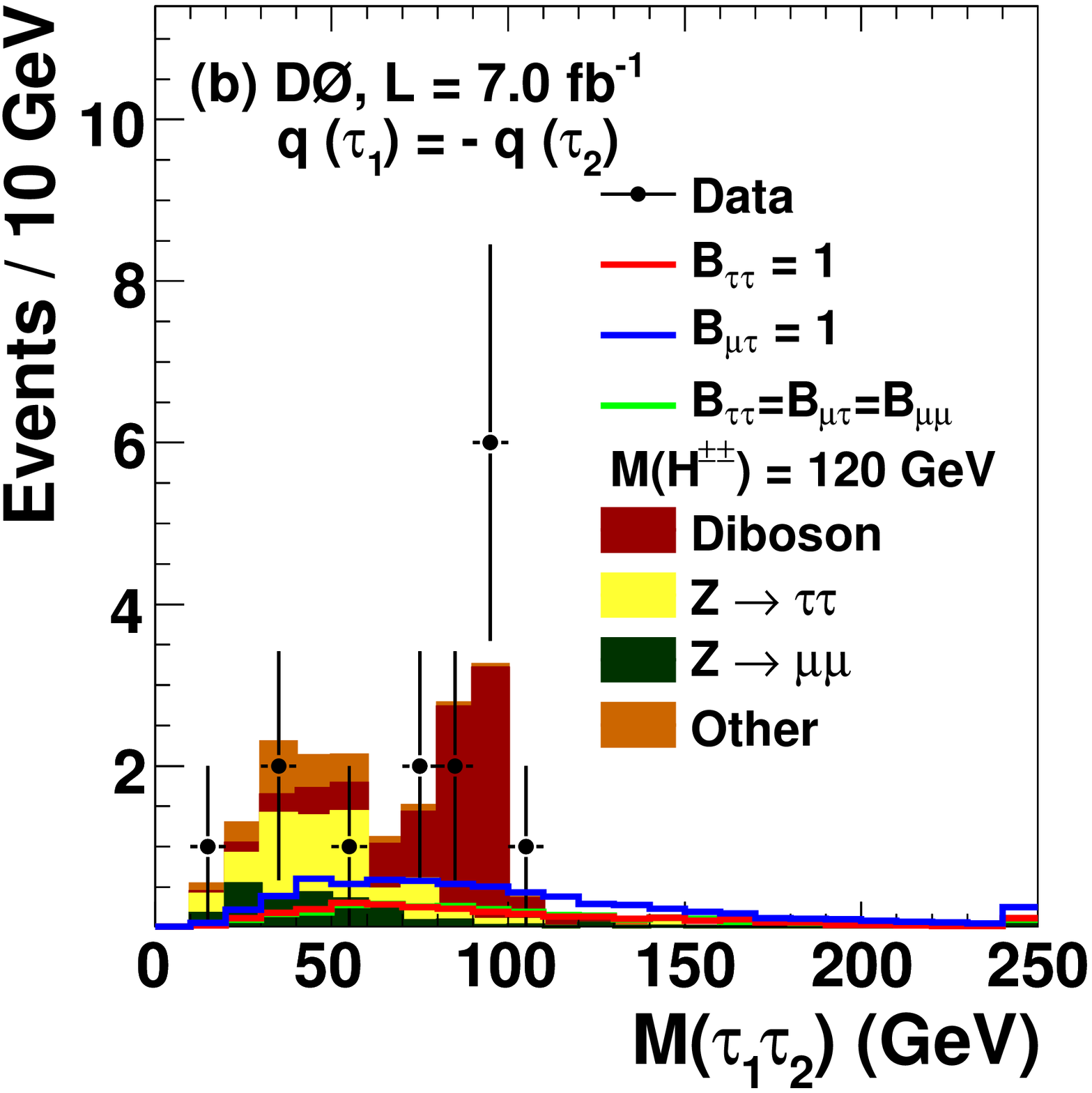}
\includegraphics[scale=0.29]{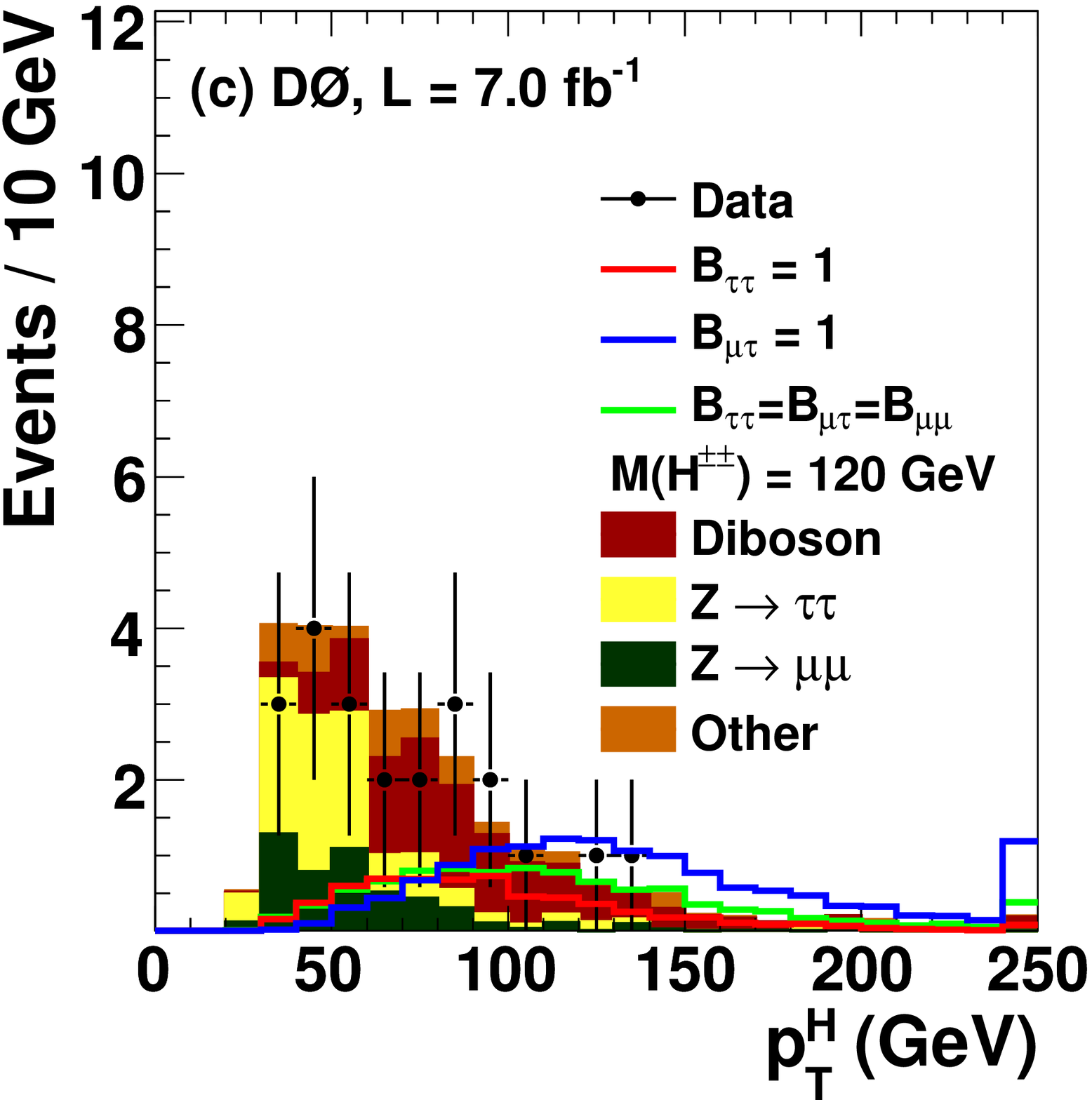}
\caption{\label{fig-kine} (color online).
$M(\tau_1,\tau_2)$ distribution 
for the (a) $q_{\tau_1}=q_{\tau_2}$ and (b) $q_{\tau_1}=-q_{\tau_2}$ samples, and (c) transverse
momentum of the doubly-charged dilepton system $p_T^H$, for all four
samples combined, after all selections. 
The data are compared to the sum of the expected background and
to simulations of a $\HpmL\HpmL$ signal for $M(\Hpm)=120$~GeV and 
$\BR(\HpmL\to\tau^{\pm}\tau^{\pm})=1$, $\BR(\HpmL\to\mu^{\pm}\tau^{\pm})=1$,  and
$\BR(\HpmL\to\tau^{\pm}\tau^{\pm})=\BR(\HpmL\to\mu^{\pm}\mu^{\pm})
     =\BR(\HpmL\to\mu^{\pm}\tau^{\pm})=1/3$, normalized
using the NLO calculation of the cross section.
``Other" background comprises $W$+jet, 
$Z/\gamma^{*}\to {e^+e^-}$, and $t\bar{t}$ processes.
All entries exceeding the range of the
     histogram are added to the last bin.
}
\end{figure*}

All background processes are simulated using Monte Carlo (MC) event generators, 
except the multijet background, which is determined from data. 
The $W$+jet, $Z/\gamma^{*}\to {\ell^+\ell^-}$, and $t\bar{t}$ processes 
are generated using {\sc alpgen}~\cite{bib-alpgen} with showering and
hadronization provided by {\sc pythia}~\cite{bib-pythia}.  Diboson production 
({\sl WW}, {\sl WZ}, and {\sl ZZ}) and 
signal events are simulated using {\sc pythia}. 
The signal samples for the model with equal branching ratios for the decays
$\Hpm\to\tau^{\pm}\tau^{\pm}$, $\mu^{\pm}\mu^{\pm}$, and
$\mu^{\pm}\tau^{\pm}$ are generated using Yukawa couplings
$h_{\mu\tau}=h_{\tau\mu}=\sqrt{2}h_{\tau\tau}=\sqrt{2}h_{\mu\mu}$.
The tau lepton decays are simulated with {\sc tauola}~\cite{bib-tauola}, 
which includes a full treatment of the tau polarization.
All MC samples are processed through a 
{\sc geant}~\cite{bib-geant} simulation of the detector. Data 
from random beam crossings are overlaid on MC events to account for detector noise and additional $\ppbar$ interactions.
The simulated  distributions are corrected for the dependence of the trigger efficiency in data on
the instantaneous luminosity and for differences between data and simulation in the reconstruction 
efficiencies and in the distribution of the longitudinal
coordinate of the interaction point along the beam direction.
Next-to-leading order (NLO) quantum chromodynamics calculations of cross sections are 
used to normalize the  
signal and the background contribution of diboson processes, 
and next-to-NLO calculations are used for all other processes.

Two types of tau lepton decays into hadrons ($\tau_h$) 
are identified by their signatures:
Type-1 tau candidates consist of a calorimeter cluster, 
with one associated track and no
subcluster in the EM section of the calorimeter. This signature corresponds mainly 
to $\tau^{\pm} \rightarrow \pi^{\pm} \nu$ decays. 
For type-2 tau candidates, an energy deposit in
the EM calorimeter is required in addition to the type-1 signature, as expected for 
$\tau^{\pm} \rightarrow \pi^{\pm} \pi^{0} \nu$ decays. 
The outputs of neural networks, one for each tau-type,
designed to discriminate $\tau_h$ from jets,
have to be {\it NN}$_{\tau}>0.75$~\cite{d0-z-tautau}.
Their input variables are based on isolation variables for objects
and on the spatial distribution of showers. 
The tau lepton energy is measured with the calorimeter.

We select events with at least one muon and at least two 
$\tau_h$ candidates.
The muons must be isolated, both in the tracking detectors and in
the calorimeters.
Each event must have a reconstructed $\ppbar$ interaction 
vertex with a longitudinal component located within $60$~cm 
of the nominal center of the detector. 
The longitudinal coordinate $z_{\rm dca}$ of the distance of closest approach 
for each track is measured with respect to the nominal center of the detector.
The differences between $z_{\rm dca}$ of
the highest-$p_T$ muon and the two highest-$p_T$ $\tau_h$
(labeled $\tau_1$ and $\tau_2$), must be less than $2$~cm.
The pseudorapidity~\cite{eta} of the selected muons, $\tau_1$, and $\tau_2$
must be $|\eta^{\mu}| < 1.6$  and $|\eta^{\tau_{1,2}}| < 1.5$, respectively, and 
for additional $\tau_h$ candidates we require $|\eta^{\tau}| < 2$.
The transverse momenta  must be
$p_T^{\mu}>15$~GeV and $p_T^{\tau_{1,2}}>12.5$~GeV.
All selected $\tau_h$ candidates and muons
are required to be separated by 
$\Delta {\cal R}_{\mu\tau} > 0.5$, where $\Delta{\cal R}=\sqrt{(\Delta\phi)^2+(\Delta\eta)^2}$
and $\phi$ is the azimuthal angle,
and the two leading $\tau_h$ must be separated by
$\Delta {\cal R}_{\tau_1\tau_2} > 0.7$.
The sum of the charges of the highest-$p_T$ muon, $\tau_1$, and $\tau_2$ 
is required to be
$Q=\sum_{i=\mu,\tau_1,\tau_2} q_i=\pm 1$ as expected for signal.
After all selections, the main background is from diboson production
and $Z\to\tau^+\tau^-$, where an additional jet mimics a lepton.

\begin{table}[htbp]
\renewcommand{\arraystretch}{1.0}
\renewcommand{\tabcolsep}{0.5mm}
\caption{Numbers of events in data, predicted background, and expected 
signal for $M(\HpmL)=120$~GeV, 
assuming the NLO calculation of the signal cross section for
 $\BR(\HpmL\to\tau^{\pm}\tau^{\pm})=1$,
 $\BR(\HpmL\to\mu^{\pm}\tau^{\pm})=1$,  and
$\BR(\HpmL\to\tau^{\pm}\tau^{\pm})=\BR(\HpmL\to\mu^{\pm}\mu^{\pm})
   =\BR(\HpmL\to\mu^{\pm}\tau^{\pm})=1/3$.
 The numbers are shown for the four samples separately, together
 with their total uncertainties. }
\begin{center}
\begin{footnotesize}
   \begin{tabular}{c|ccccc}
   \hline\hline
& All &  \multicolumn{2}{c}{$N_{\mu}=1$} &
     $N_{\mu}=1$  & $N_{\mu}=2$ \\
 &  &  \multicolumn{2}{c}{$N_{\tau}=2$} 
 &  $N_{\tau}= 3$  & $N_{\tau}=2$ \\
 &  & $q_{\tau_1}=q_{\tau_2}$   &  $q_{\tau_1}=-q_{\tau_2}$ &  &  \\
       \hline
  Signal &&&&&\\
$\tau^{\pm}\tau^{\pm}$&  $6.6 \pm 0.9$  & $1.4 \pm 0.2$ & $3.1 \pm 0.4$  & $1.6\pm 0.2$ & $0.4\pm 0.1$  \\ 
$\mu^{\pm}\tau^{\pm}$	& $13.9 \pm 1.9\phantom{0}$ & $0.3 \pm 0.1$ & $6.8 \pm 0.9$  & $0.4\pm 0.1$  & $6.3\pm 0.9$  \\
Equal $\BR$ & $9.5 \pm 1.3$ & $2.5 \pm 0.3$ & $3.1 \pm  1.0$ & $1.2\pm 0.2$  & $2.6\pm 0.4$  \\
    \hline
           Background &&&&&\\
$Z\to\tau^+\tau^-$  & $8.2 \pm 1.1$	& $3.4 \pm 0.5$  & $4.8 \pm 0.7$ & $<0.1$ & $<0.1$  \\
$Z\to \mu^+\mu^-$ & $5.1 \pm 0.7$	& $2.2 \pm 0.3$  & $2.5 \pm 0.4$ & $0.1 \pm 0.1$ & $0.2 \pm   0.1$  \\
$Z\to e^+e^-$         & $0.3 \pm 0.1$	&  $<0.1$  	  & $0.3 \pm 0.1$ & $<0.1$              & $<0.1$   \\ 
$W$ + jets       & $2.9 \pm 0.4$ &  $1.1 \pm 0.2$ & $1.8 \pm 0.3$ & $<0.1$              & $<0.1$  \\
$t\bar{t}$         & $0.6 \pm 0.1$	&  $0.3 \pm 0.1$ & $0.3 \pm 0.1$ & $0.1 \pm 0.1$   & $<0.1$  \\ 
Diboson	       & $10.5 \pm 1.7$& $0.5 \pm 0.1$ & $8.5  \pm 1.4$& $0.4 \pm 0.1$   & $1.1\pm0.2$   \\
Multijet            & $<0.8$ & $<0.2$ & $<0.5$ & $<0.1$ & $<0.1$ \\
   \hline
   Background & & & & &\\
   Sum & $27.6 \pm 4.9$ & $7.5 \pm 1.2$ & $18.2 \pm 3.3$ & $0.6 \pm 0.1$  & $1.3 \pm 0.2$ \\
\hline
         Data       &    $22$  &  $5$   &  $15$ &  $0$ &  $2$     \\ 
 \hline\hline
\end{tabular}
\label{tab-events}
\end{footnotesize}
\end{center}
\end{table}

We estimate the multijet background using three independent data 
samples and identical selections, except with the {\it NN}$_{\tau}$ requirements  reversed, by requiring that either one or both $\tau_h$ candidates 
have {\it NN}$_{\tau}<0.75$. The simulated background
is subtracted before the samples are used to determine the differential distributions
and normalization of the multijet background in the signal region. A second
method used to estimate the multijet background 
is based on the fact that events with $Q = \pm 1$ are signal-like, 
whereas events with 
$Q=\pm 3$ correspond largely to multijet background. To reduce
the $W$+jets contribution in the sample with $Q=\pm 3$,
the visible $W$ boson mass  
$M_W = \sqrt{2p^{\mu}\MET(1-\cos\phi)}$ is required to be
$<50$~GeV, where $p^{\mu}$ is the muon momentum, $\MET$ the imbalance
in transverse momentum measured in the calorimeter, and
$\phi$ is the azimuthal angle between the muon and the direction of the $\MET$. 
The total rate of expected multijet background events following all selections 
is negligible ($<3\%$ of the total background). 
We also use the sample where both $\tau_h$ candidates 
have {\it NN}$_{\tau}<0.75$ to study the rate of jets that are falsely reconstructed
as $\tau_h$ and we find this rate to be well modeled by the simulation.

\begin{figure}[htbp]
   \begin{center}	      
     \includegraphics[width=0.42\textwidth]{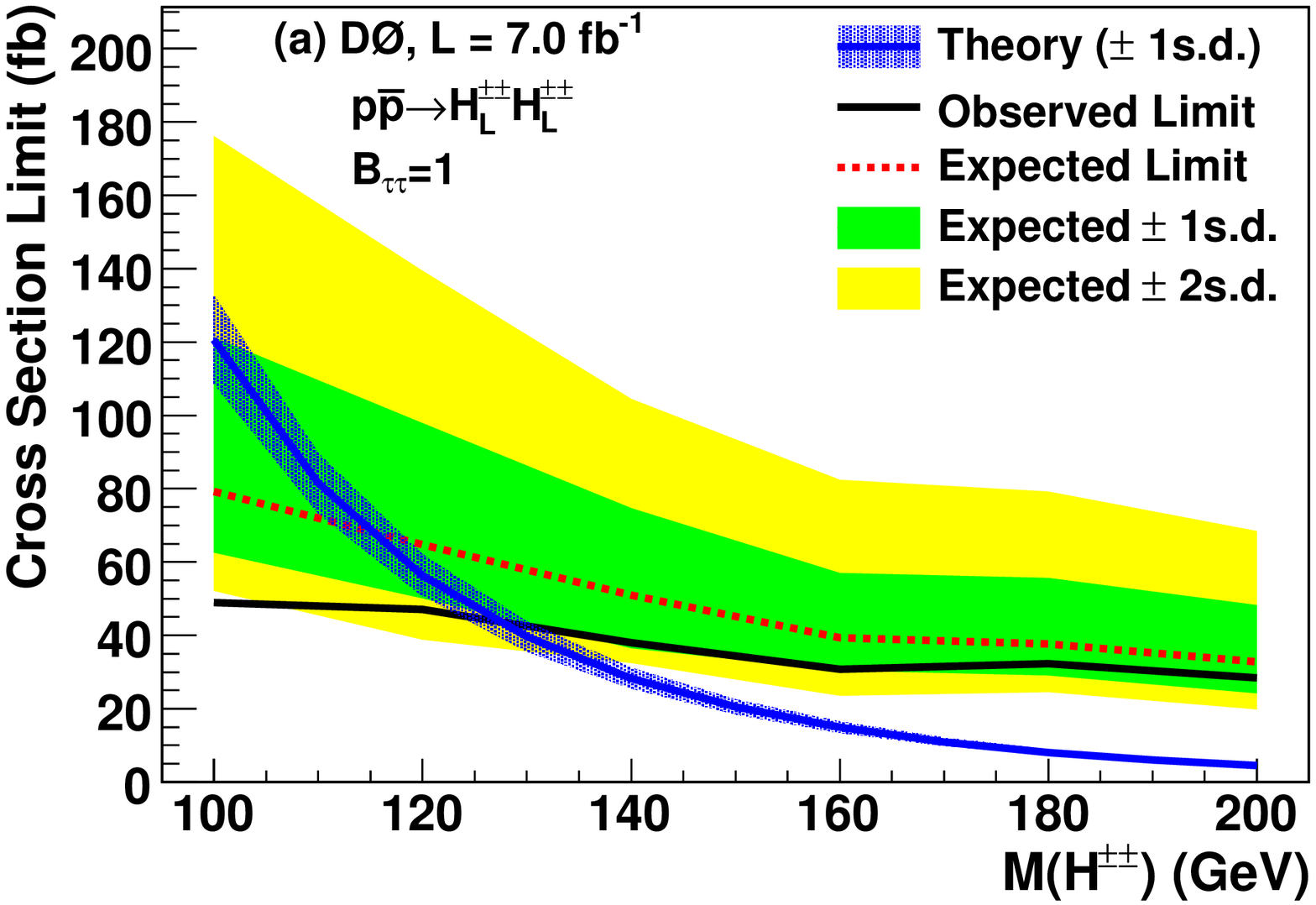}\\
     \includegraphics[width=0.42\textwidth]{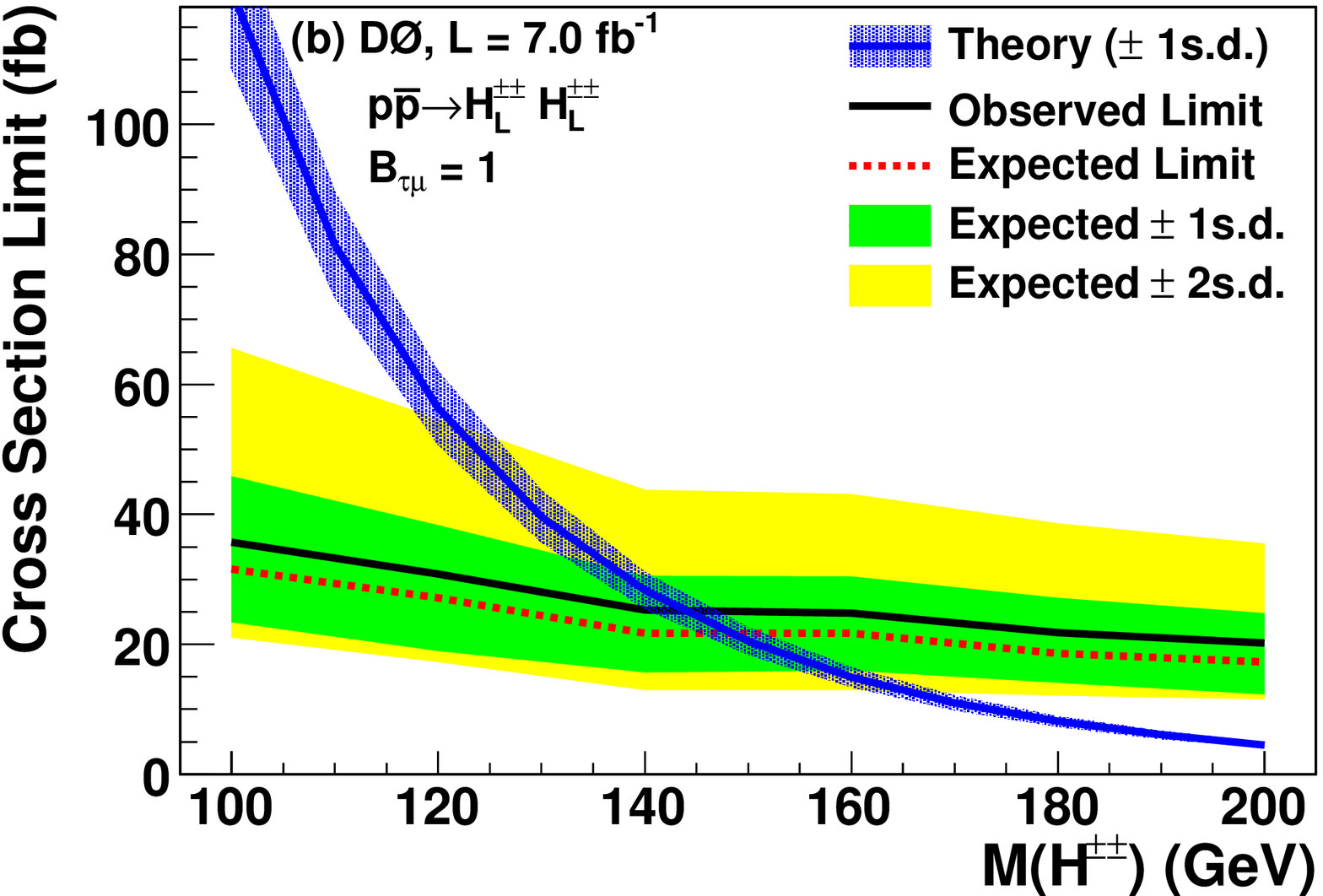}\\
     \includegraphics[width=0.42\textwidth]{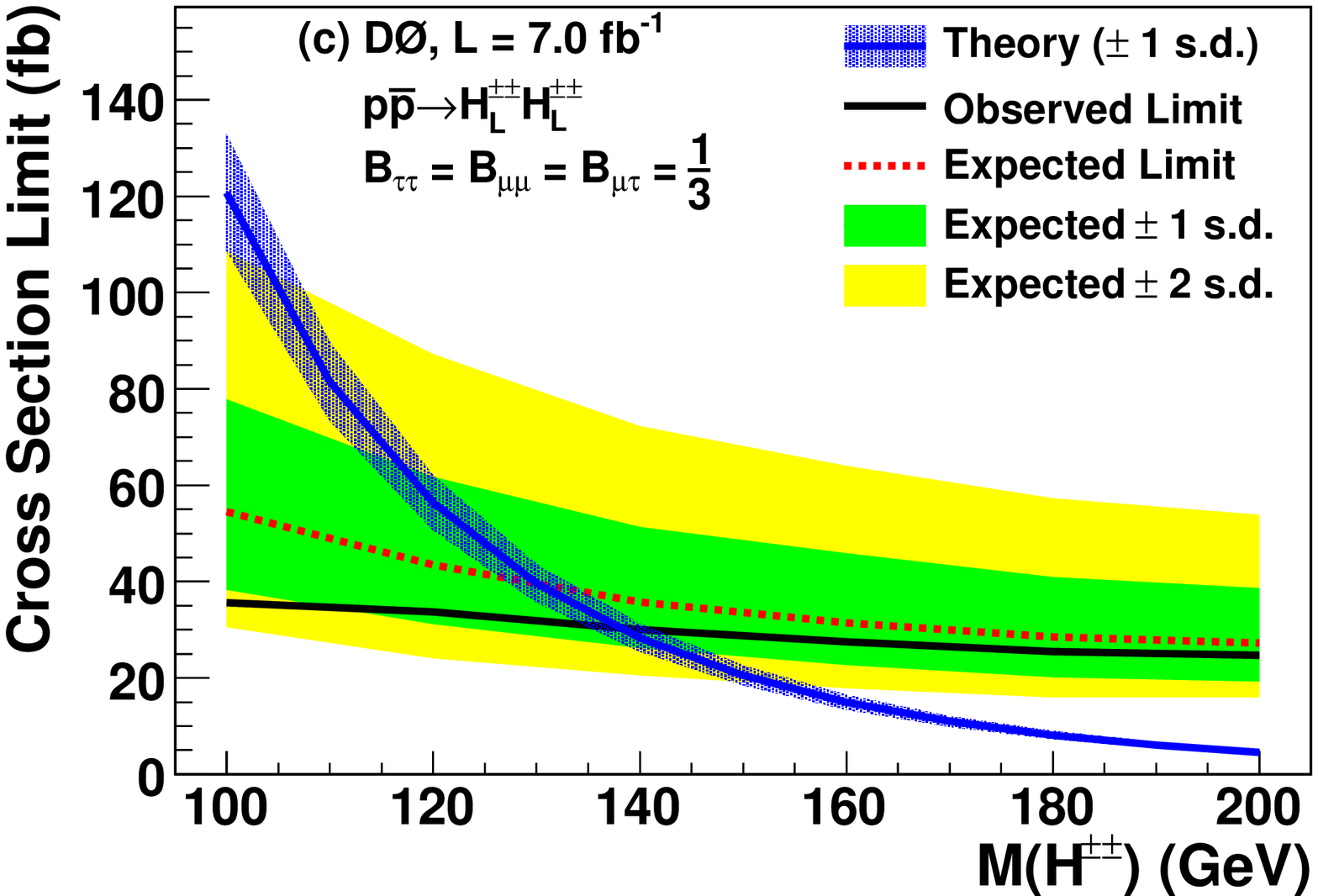}
   \caption{(color online).
     Upper limit on the $\HpmL\HpmL$ pair production cross section for 
     (a) $\BR(\HpmL\to\tau^{\pm}\tau^{\pm})=1$,
     (b) $\BR(\HpmL\to\mu^{\pm}\tau^{\pm})=1$,  and
     (c) $\BR(\HpmL\to\tau^{\pm}\tau^{\pm})=\BR(\HpmL\to\mu^{\pm}\mu^{\pm})
     =\BR(\HpmL\to\mu^{\pm}\tau^{\pm})=1/3$.
     The bands around the median expected limits correspond to regions of $\pm 1$ and $\pm 2$ standard deviation (s.d.), and
    the band around the predicted NLO cross section for signal corresponds to a
     theoretical uncertainty
     of $\pm 10\%$.
    }
\label{fig-limits}
\end{center}
\end{figure}

To improve the discrimination of signal from background, 
the data are subdivided into
four nonoverlapping samples, depending on the
charges of the muon ($q_{\mu}$) and the $\tau_h$ candidates ($q_{\tau}$)
and the number of muons ($N_{\mu}$) and $\tau_h$ 
($N_{\tau}$) in the event. 
First, we define two samples for events 
with $N_{\mu}=1$ and $N_{\tau}= 2$. Because the two like-charge
leptons are assumed to originate from a single $\Hpm$ decay, we 
consider separately events where both tau leptons have the same charge, 
$q_{\tau_1}=q_{\tau_2}$, 
and events with $\tau_1$ and $\tau_2$ of opposite charge, i.e., 
$q_{\tau_1}=-q_{\tau_2}$, which implies that one of the $\tau$ leptons
and the muon have the same charge.
The third sample is defined by $N_{\tau}=3$ and the fourth sample by $N_{\mu}= 2$,
without any additional requirements on the charges.

The distributions of the invariant mass of the two leading tau candidates, 
$M(\tau_1,\tau_2)$,
for the like and opposite-charge samples 
are shown in Figs.~\ref{fig-kine}(a) and (b).
The separation into samples with different fractions
of signal and background events increases the sensitivity to signal, as the composition of 
the background is different, with 
the like-charge sample being dominated by background from  $Z$+jets 
decays and the opposite-charge sample by background from
diboson production. The diboson background is mainly due to {\sl WZ}~$\to\mu\nu e^+e^-$ events
where the electrons are misidentified as tau leptons.
In Fig.~\ref{fig-kine}(c) we show the transverse momentum of the doubly-charged
dilepton system, $p_T^H$, which corresponds to the reconstructed 
$\Hpm\to\ell^{\pm}\ell'^{\pm}$ decay, where $\ell^{\pm}\ell'^{\pm}=
(\mu^{\pm}\tau_1^{\pm},\mu^{\pm}\tau_2^{\pm},\tau_1^{\pm}\tau_2^{\pm})$ is the pairing of the two highest-$p_T$ $\tau_h$ and the highest-$p_T$ muon that have the same charges.
Since $|Q|=1$, only one such pairing exists per event.
The expected number of background and signal events for the four samples
and the observed numbers of events in data are shown in Table~\ref{tab-events}
with the statistical uncertainties of the MC samples and systematic uncertainties added in quadrature.

\begin{table}[htbp]
 \renewcommand{\tabcolsep}{0.5mm}
\caption{Expected and observed limits on
$M(\Hpm)$ (in GeV) 
for left-handed and right-handed $\Hpm$ bosons.
Only left-handed states exist in the model
that assumes equality of branching fractions into
$\tau\tau$, $\mu\tau$, and $\mu\mu$ final states. 
We only derive limits if the expected limit on $M(\Hpm)$ is $\ge 90$~GeV.
  }
 \begin{center}
   \begin{tabular}{c|cccc}
\hline\hline
   Decay & \multicolumn{2}{c}{$\HpmL$}& \multicolumn{2}{c}{$\HpmR$} \\
  \cline{2-5}
   & expected & observed & expected & observed \\
${\cal B}(\Hpm\to\tau^{\pm}\tau^{\pm})=1$ & $116$ & $128$ &   \\
${\cal B}(\Hpm\to\mu^{\pm}\tau^{\pm})=1$ & $149$ & $144$ & $119$ & $113$ \\
Equal $\BR$ into  &  & &  & \\
$\tau^{\pm}\tau^{\pm},\mu^{\pm}\mu^{\pm},\mu^{\pm}\tau^{\pm}$ & $130$ & $138$ & 
&\\
\hline 
${\cal B}(\Hpm\to\mu^{\pm}\mu^{\pm})=1$ & $180$ & $168$ & $154$ & $145$ \\
\hline\hline
 \end{tabular}
 \label{tab-limits}
 \end{center}
\end{table}

Since the data are well described by the background expectation,
we determine limits on the $\Hpp\Hmm$ production cross section 
using a modified frequentist approach~\cite{bib-wade}.  
A log-likelihood ratio (LLR) test statistic 
is formed using the Poisson probabilities for estimated
background yields, the signal acceptance, and the observed
number of events for different $\Hpm$ mass hypotheses.
The confidence levels are derived by integrating the
LLR distribution in pseudoexperiments using both the signal-plus-background 
({\it CL}$_{s+b}$) and the background-only 
hypotheses ({\it CL}$_b$). The excluded production cross section is
taken to be the cross section for which the confidence
level for signal, {\it CL}$_s=${\it CL}$_{s+b}/${\it CL}$_b$, equals $0.05$. 
The $M(\tau_1,\tau_2)$ distribution is used to discriminate signal from
background.

Systematic uncertainties on both background and signal, 
including their correlations, are taken into account.
The theoretical uncertainty on background cross sections for 
$Z/\gamma^{*}\to\ell^+\ell^-$, 
$W$+jets, $t\bar{t}$, and dibo son production vary between $6\%-10\%$.
The uncertainty on the measured integrated luminosity is $6.1\%$~\cite{bib-lumi}.
The systematic uncertainty on muon identification is $2.9\%$ per muon
and the uncertainty on the identification of $\tau_h$, including the uncertainty
from applying a neural network to discriminate $\tau_h$ from jets, is $4\%$ for
each type-1 and $7\%$ for each type-2 $\tau_h$ candidate. The trigger
efficiency has a systematic uncertainty of $5\%$. The uncertainty
on the signal acceptance from parton distribution functions is~$4\%$.

In Fig.~\ref{fig-limits}, the upper limits 
on the cross sections are compared to the NLO signal cross sections
for $\HpmL\HpmL$ pair production~\cite{bib-spira}
for some of the branching ratios considered. 
The corresponding expected and observed limits are shown in Table~\ref{tab-limits}. 

\begin{figure}[tb]
   \begin{center}	       
 \includegraphics[width=0.42\textwidth]{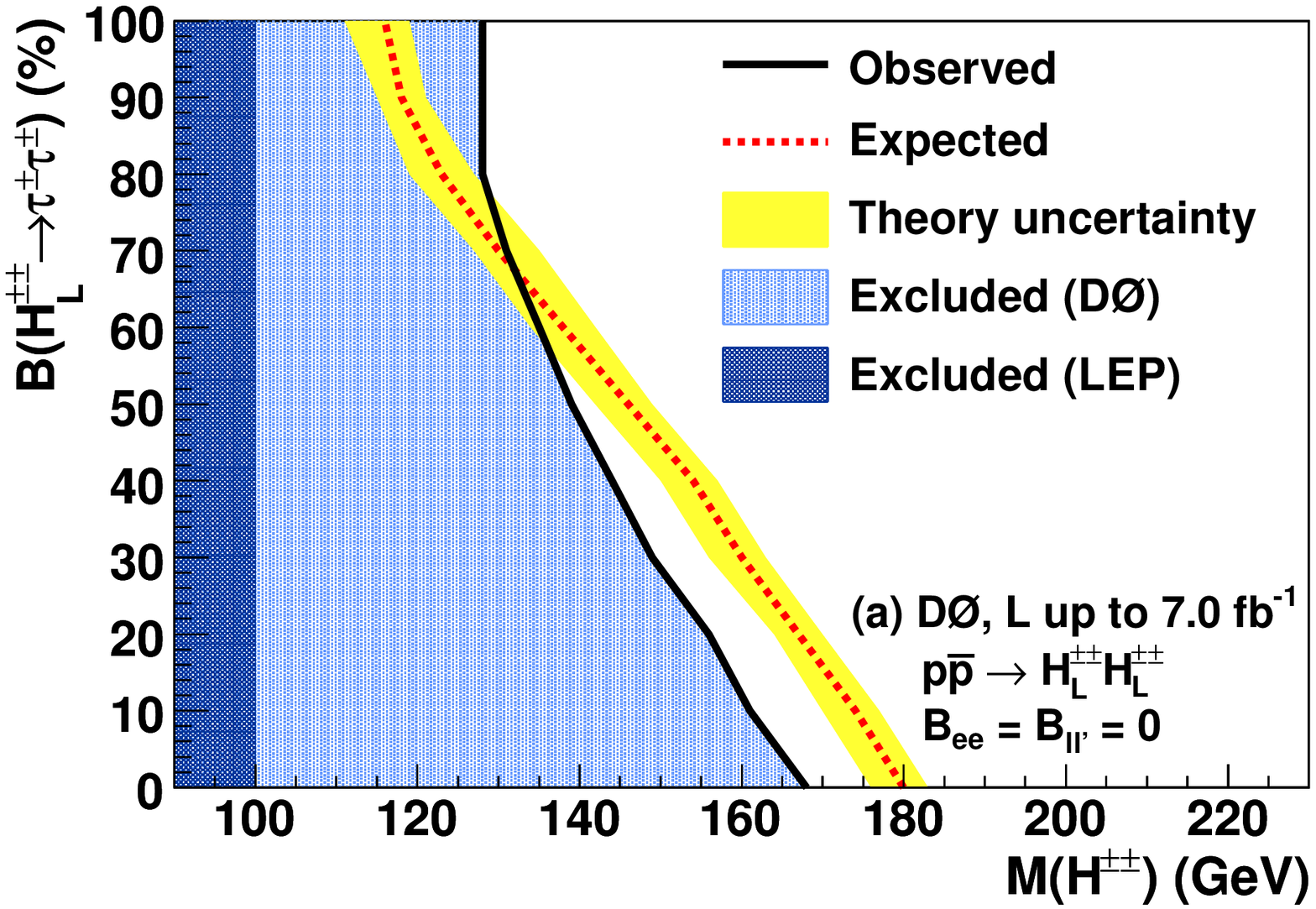}
 \includegraphics[width=0.42\textwidth]{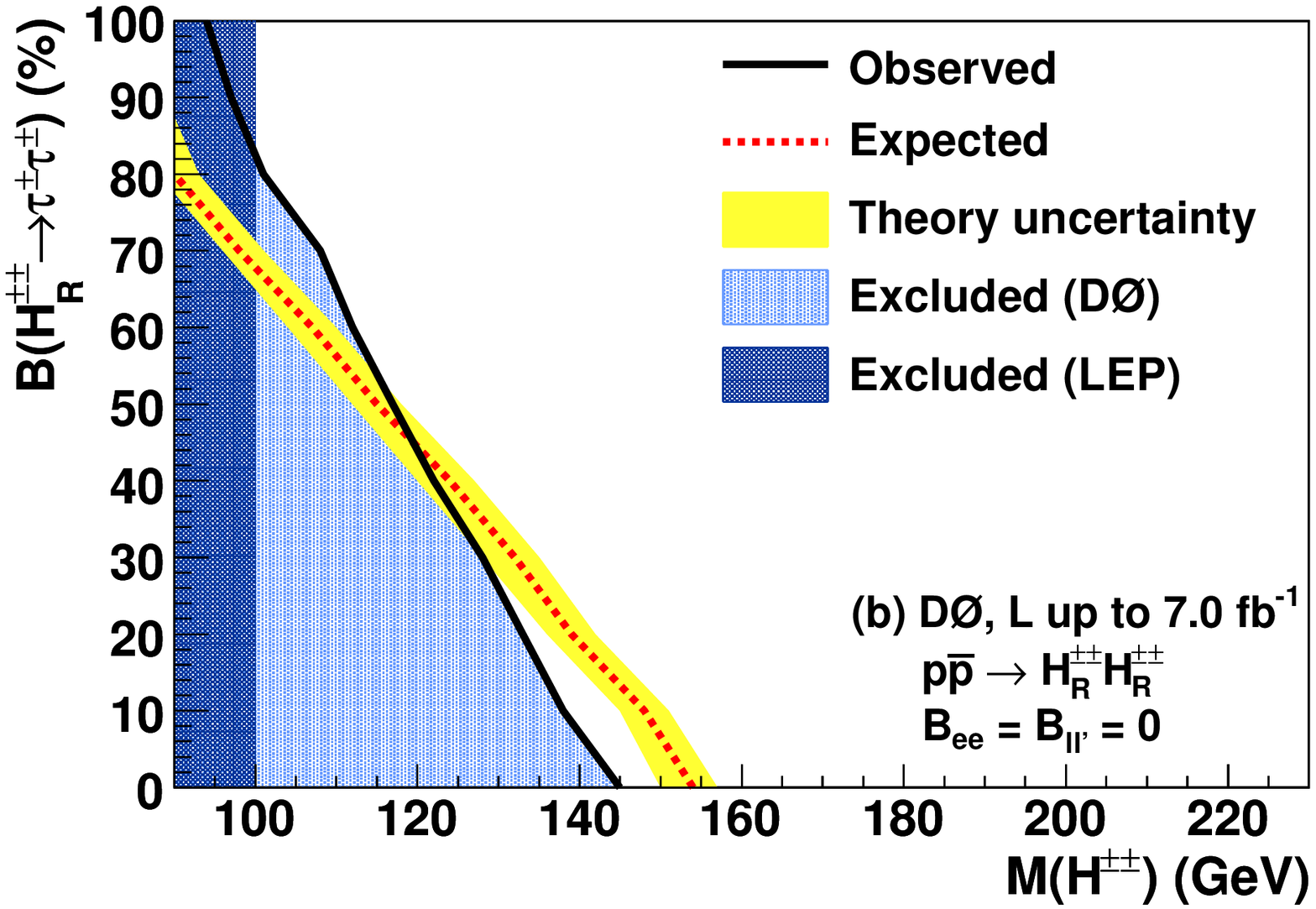}
   \caption{(color online).
    Expected and observed exclusion region at the $95\%$ C.L. in
    the plane of $\BR(\Hpm\to\tau^{\pm}\tau^{\pm})$ versus $M(\Hpm)$, assuming 
    $\BR(\Hpm\to\tau^{\pm}\tau^{\pm})+\BR(\Hpm\to\mu^{\pm}\mu^{\pm})=1$, for
     (a) left-handed and (b) right-handed $\Hpm$ bosons. The band
    around the expected limit represents the uncertainty on the 
    NLO calculation of the cross section for signal.
    }
\label{fig-result}
\end{center}
\end{figure}

The $\Hpm$ boson mass limits assuming 
$\BR(\Hpm\to\tau^{\pm}\tau^{\pm})+\BR(\Hpm\to\mu^{\pm}\mu^{\pm})=1$ are 
determined by combining signal samples generated
with pure $4\tau$, $(2\tau/2\mu)$, and $4\mu$ 
final states
with fractions $\BR^2$, $2\BR(1-\BR)$, and $(1-\BR)^2$, respectively,
where $\BR\equiv\BR(\Hpm\to\tau^{\pm}\tau^{\pm})$. 
Here, we include in the limit setting the distribution
of the invariant mass of the two highest $p_T$ muons,
including the systematic uncertainties and their correlations, 
from a search for $\Hpp\Hmm\to 4\mu$ decays performed by
the D0 Collaboration in $1.1$~fb$^{-1}$ of integrated luminosity~\cite{bib-d0hmm1}.
The results are shown
in Fig.~\ref{fig-result} for varying $\BR=0\%-100\%$ in steps
of $10\%$. 
When performing this analysis, we found that
the statistical uncertainties on the background simulations were overestimated 
in~\cite{bib-d0hmm1}. A standard treatment of the uncertainties in the limit
setting improves the mass limits
for the $4\mu$ final state, as shown in Table~\ref{tab-limits}.

In summary, we have performed the first search at a hadron collider 
for pair production of
doubly-charged Higgs bosons decaying exclusively into tau leptons. 
We set an observed (expected) lower limit of $M(\HpmL) > 128~(116)$~GeV for a
$100\%$ branching fraction of $\Hpm\to\tau^{\pm}\tau^{\pm}$, 
$M(\HpmL) > 144~(149)$~GeV for a
$100\%$ branching fraction into $\mu\tau$, and $M(\HpmL) > 130~(138)$~GeV
for a model with equal branching ratios into $\tau\tau$, $\mu\tau$,
and $\mu\mu$.  These are the most stringent limits on $\Hpm$ boson masses in these decay channels.

%
We thank the staffs at Fermilab and collaborating institutions,
and acknowledge support from the
DOE and NSF (USA);
CEA and CNRS/IN2P3 (France);
FASI, Rosatom and RFBR (Russia);
CNPq, FAPERJ, FAPESP and FUNDUNESP (Brazil);
DAE and DST (India);
Colciencias (Colombia);
CONACyT (Mexico);
KRF and KOSEF (Korea);
CONICET and UBACyT (Argentina);
FOM (The Netherlands);
STFC and the Royal Society (United Kingdom);
MSMT and GACR (Czech Republic);
CRC Program and NSERC (Canada);
BMBF and DFG (Germany);
SFI (Ireland);
The Swedish Research Council (Sweden);
and
CAS and CNSF (China).


\begin{thebibliography}{99}

\bibitem{bib-littleH}
N. Arkani-Hamed {\sl et al.}, J. High Energy Phys. {\bf 08}, 021 (2002).

\bibitem{bib-LRsym}
 R.N.~Mohapatra and G.~Senjanovic, Phys.\ Rev.\ Lett.\  {44}, 912 (1980);
 R.N.~Mohapatra and G.~Senjanovic, Phys.\ Rev.\  {\bf D 23}, 165 (1981);
J.F.~Gunion {\sl et al.}, Phys. Rev. D {\bf 40}, 1546 (1989);
N.G.~Deshpande {\sl et al.}, Phys. Rev. D {\bf 44}, 837 (1991). 

\bibitem{bib-331}
J.E.~Cieza Montalvo {\sl et al.}, 
Nucl. Phys. {\bf B756}, 1 (2006); erratum-ibid. {\bf B796}, 422 (2008).

\bibitem{bib-single}
A.G.~Akeroyd and M.~Aoki, Phys. Rev. D {\bf 72}, 035011 (2005).

\bibitem{bib-mass}
C.S.~Aulakh, A.~Melfo, and G.~Senjanovic, Phys. Rev. D {\bf 57}, 4174 (1998);
Z.~Chacko and R. N.~Mohapatra, Phys. Rev. D {\bf 58}, 015003 (1998).
 
\bibitem{bib-scalar}
E.~Ramirez Barreto, Y.A.~Coutinho, and J.~S\'a Borges, Phys. Rev. D {\bf 83}, 075001 (2011).

\bibitem{bib-seesaw}
M. Kadastik, M. Raidal, and L. Rebane, Phys. Rev. D {\bf 77}, 115023 (2008);
A. Hektor {\sl et al.}, 
Nucl. Phys. {\bf B787}, 198 (2007).
 
\bibitem{bib-spira}
M.~M\"uhlleitner and M.~Spira,  Phys.~Rev.~D {\bf 68}, 117701 (2003) and private communications. 

\bibitem{bib-lep}
G.~Abbiendi {\sl et al.} (OPAL Collaboration), Phys. Lett. B {\bf 526}, 221 (2002);
P.~Achard {\sl et al.} (L3 Collaboration), Phys. Lett. B {\bf 576}, 18 (2003);
J.~Abdallah {\sl et al.} (DELPHI Collaboration), Phys. Lett. B {\bf 552}, 127 (2003).

\bibitem{bib-hera}
A.~Aktas {\sl et al.} (H1 Collaboration), Phys.~Lett.~B {\bf 638}, 432 (2006). 

\bibitem{bib-opal}
G.~Abbiendi {\sl et al.} (OPAL Collaboration), Phys. Lett. B {\bf 577}, 93 (2003).

\bibitem{bib-moha1}
J.F.~Gunion {\sl et al.}, Phys. Rev. D {\bf 40}, 1546 (1989); 
R.N.~Mohapatra, Phys. Rev. D {\bf 46}, 2990 (1992).

\bibitem{bib-d0hmm1}
V.M.~Abazov {\sl et al.} (D0 Collaboration), Phys. Rev. Lett. {\bf 101}, 071803 (2008).

\bibitem{bib-d0hmm2}
V.M.~Abazov {\sl et al.} (D0 Collaboration), Phys. Rev. Lett. {\bf 93} 141801 (2004).

\bibitem{bib-cdf1}
D.~Acosta {\sl et al.} (CDF Collaboration), Phys. Rev. Lett. {\bf 93}, 221802 (2004).

\bibitem{bib-cdf2}
T.~Aaltonen {\sl et al.} (CDF Collaboration), Phys. Rev. Lett. {\bf 101}, 121801 (2008).

\bibitem{d0det}
V.M.~Abazov {\sl et al.} (D0 Collaboration),
Nucl. Instrum. Methods Phys. Res. A {\bf 565}, 463  (2006);
M.~Abolins {\sl et al.}, Nucl. Instrum. Methods in Phys. Res.
A {\bf 584}, 75 (2008);
R.~Angstadt {\sl et al.}, Nucl.~Instrum.~Methods in Phys. Res.
A {\bf 622}, 298 (2010).

\bibitem{bib-alpgen}
M.L.~Mangano {\sl et al.}, J. High Energy Phys. {\bf 07}, 1 (2003);
we use version 2.11.
 
\bibitem{bib-pythia}
T.~Sj\"ostrand, S. Mrenna, and P. Skands, 
J. High Energy Phys. {\bf 05}, 026 (2006); we use version 6.323.

\bibitem{bib-tauola}
Z.~Wa\c{s}, Nucl. Phys. {\bf B} Proc. Suppl. {\bf 98}, 96 (2001);  we use version 2.5.04.

\bibitem{bib-geant}
R.~Brun and F. Carminati, CERN Program Library Long Writeup W5013, 1993.

\bibitem{d0-z-tautau} 
V.M.~Abazov {\sl et al.} (D0 Collaboration), Phys. Rev. D {\bf 71}, 072004 (2005);
erratum-ibid.\  D {\bf 77},  039901 (2008).

\bibitem{eta}
The pseudorapidity is defined as $\eta = -\ln[\tan(\theta/2)]$, where $\theta$ is the polar angle with respect to the proton beam direction.

\bibitem{bib-wade}
W.~Fisher, FERMILAB-TM-2386-E (2006).
 
 \bibitem{bib-lumi}
T.~Andeen {\sl et al.}, FERMILAB-TM-2365 (2007).

\end{thebibliography}
\end{document}